\documentclass[authoryear,preprint,10pt,3p]{elsarticle}

\usepackage{comment}
\usepackage{float}
\usepackage{ulem}
\usepackage{amssymb}
\usepackage{tikz}
\usepackage{mathtools}
\usepackage{natbib}
\usepackage{multirow}
\usepackage{color}
\usepackage{soul}
\usepackage{graphicx}
\usepackage{caption}
\usepackage{xcolor}
\usepackage{algorithm}
\usepackage{algorithmicx}
\usepackage{algpseudocode}
\usepackage[unicode=true]{hyperref}
\let\pkg=\texttt

\usepackage{booktabs}

\usepackage{longtable}
\usepackage[edges]{forest}

\journal{International Journal of Forecasting}
\begin{document}
\begin{frontmatter}

\title{Exploring the representativeness of the M5 competition data}

\author[label1]{Evangelos Theodorou \fnref{eqc}}
\ead{vagtheodorou@fsu.gr}
\author[label2]{Shengjie Wang \fnref{eqc}}
\ead{wsj19992017@buaa.edu.cn}
\author[label2]{Yanfei Kang\corref{cor1}}
\ead{yanfeikang@buaa.edu.cn}
\cortext[cor1]{Corresponding author}
\author[label1]{Evangelos Spiliotis}
\ead{spiliotis@fsu.gr}
\author[label3]{Spyros Makridakis}
\ead{makridakis.s@unic.ac.cy}
\address[label1]{Forecasting and Strategy Unit, School of Electrical and Computer Engineering, National Technical University of Athens, Greece}
\address[label2]{School of Economics and Management, Beihang University, China}
\address[label3]{Institute For the Future, University of Nicosia, Cyprus}
\author[label1]{Vassilios Assimakopoulos}
\ead{vassim@fsu.gr}
\fntext[eqc]{The authors contributed equally. }

\begin{abstract}
The main objective of the M5 competition, which focused on forecasting the hierarchical unit sales of Walmart, was to evaluate the accuracy and uncertainty of forecasting methods in the field in order to identify best practices and highlight their practical implications. However, whether the findings of the M5 competition can be generalized and exploited by retail firms to better support their decisions and operation depends on the extent to which the M5 data is sufficiently similar to unit sales data of retailers that operate in different regions, sell different types of products, and consider different marketing strategies. To answer this question, we analyze the characteristics of the M5 time series and compare them with those of two grocery retailers, namely Corporación Favorita and a major Greek supermarket chain, using feature spaces. Our results suggest that there are only small discrepancies between the examined data sets, supporting the representativeness of the M5 data.
\end{abstract}

\end{frontmatter}

\newpage
\section{Introduction}

Time series forecasting competitions provide valuable insights when it comes to identifying the most appropriate methods for producing point or probabilistic forecasts for a forecasting task under investigation, with their findings having several implications, both for the industry and the academia \citep{HYNDMAN20207}. This is especially true when the time series data considered by the competitions and the forecasting methods originally submitted are publicly available to use, enabling the replication of their results \citep{MAKRIDAKIS2018}. This was the case with M5, which focused on forecasting the hierarchical unit sales of the largest retail company in the world, Walmart \citep{M5special_data}.

The results of the M5 competition demonstrate that there is still room for improving forecasting accuracy and uncertainty estimation in the retail industry. Similar to M4 \citep{MAKRIDAKIS202054}, traditional statistical methods, widely used by retailers for supporting decisions related to supply chain management, were outperformed by state-of-the-art machine learning ones, with the improvements reported for the winning submissions over the top-performing benchmarks being higher than 20\% according to the competition's evaluation measures. This finding indicates that retail and logistic firms could benefit significantly from utilizing the winning submissions of the competition, especially if we consider that small improvements in accuracy can lead to considerable inventory reductions \citep{SYNTETOS2010134}, while slight inaccuracies to higher stock holdings and lower service levels \citep{GHOBBAR20032097, Pooya2019}.

However, the practical implications of forecasting competitions have been widely criticized by the forecasting community and especially its practitioners, claiming that the findings reported may depend on the particularities of the data set used for conducting the competition, thus being difficult to generalize and exploit in business, real-life applications \citep{ORD2001,CLEMENTS2001,ARMSTRONG2019,DARIN2019,FRY2019,BOJER2021587}. For example, M5 focused on the sales of ten indicative US stores of a global retail firm located in three states (California, Wisconsin, and Texas), covering the period from 2011 to 2016 and three product categories (``Foods'', ``Household'', and ``Hobbies''). Therefore, it could be the case that its results are not representative of a retail firm operating in South America in the same period or another retailer operating in Greece in a different period. Indeed, the country of origin of the data may affect the characteristics of the series, especially if the types of the products sold and the marketing strategies considered (e.g., discounts and promotions) differ \citep{makfuture2021}. Moreover, as COVID-19 has recently proved, the period of analysis may highly affect the patterns of the data and, as a result, the appropriateness of the forecasting methods that should be employed for producing accurate forecasts \citep{WANG20202903, PANZONE2021100495, GUNGOR2021120637}. 

As the literature suggests, no single method is suitable for all forecasting tasks \citep{LAWRENCE2001} and, therefore, identifying ``horses for courses'' \citep{PETROPOULOS2014152} is essential. This was also true in the M5 competition where depending on the nature of the series, the cross-sectional level being forecast, and the quantile considered for estimating uncertainty, different methods were found to be the top-performing ones \citep{M5special_accuracy, M5special_uncertainty}. It becomes evident that although the winning method of a competition may not always outperform the rest of the available alternatives in all subsets of the competition's data set, its findings can still be useful when forecasters are able to identify and select the methods that performed best in a subset of series which represent their own data set adequately, thus fitting their forecasting needs \citep{FILDES2020186, MAKRIDAKIS2020217}. Drawing from the above, evaluating the extent to which the M5 data reflects the particularities of various retail firms becomes critical.

In this discussion paper, we try to answer this question by analyzing the time series features of the M5 data set and comparing them with those of two other retail firms, namely Corporación Favorita, a major grocery retailer in South America, and a major supermarket chain in Greece\footnote{For reasons of confidentiality, the name of the company is not provided.}. We base our analysis on feature space visualizations, as proposed by \citep{KANG2020}, and consider intuitive measures of coverage and miscoverage to quantify our results. 

We should clarify that our analysis focuses on the sales data itself, ignoring the explanatory variables that typically accompany such data to enhance forecasting performance \citep{FILDES2019R}. Therefore, the representatives of the M5 data set is evaluated based on the patterns that the M5 sales data display per se compared to that of other retail firms and not in terms of the external information that may be available to forecasters to help them explain data variations or determine changes in their future behavior. For instance, the M5 data set included information about special days and holidays, selling prices, and promotion activities. Although access in that kind of information may be typical in retail sales forecasting applications, other data sets may be richer in terms of explanatory variables, including also information about future weather conditions, product reviews, trends, and fuel prices, among others \citep{MA2021111}.

\section{Time series feature extraction and selection}

Time series feature representation approaches allow the extensive investigation and intuitive visualization of large data sets. Several studies have exploited feature-based time series instance space analysis to explore previous forecasting competition data sets~\citep[e.g.,][]{Kang2017,SPILIOTIS202037,Li2020imaging}. However, a significant difference of M5 over previous forecasting competitions is that it involved intermittent demand series that display zeros and irregular patterns. In this respect, in this study, feature extraction and selection are performed using a methodological approach tailored for the examined application, as presented in the remainder of this section.

\subsection{Feature extraction}
\subsubsection{Determining the pool of features} 

Feature extraction begins by constructing a pool of features $\mathcal{F}$, specifically by concatenating two sets of time series features, namely $\mathcal{F}_\text{tsfresh}$, from the \pkg{tsfresh} module for python \citep{Christ2018}, and  $\mathcal{F}_\text{tsfeatures}$, from the \pkg{tsfeatures} package for R \citep{Hyndman2019a}. In total, $\mathcal{F}$ consists of $P = 836$ features, computed for all the series under examination. The \pkg{tsfresh} package involves a variety of features, such as basic time series statistics, correlation measures, entropy estimations, and coefficients of standard time series forecasting and analysis methods. On the other hand, the \pkg{tsfeatures} package includes statistics that are computed on the first and second-order differences of the raw series, account for seasonality, and exploit the outputs of popular time series decomposition methods, among others.

\subsubsection{Expanding the pool of features for higher temporal aggregation levels} 

Due to the intermittency of typical retail sales series, some features like seasonality and trend may be observed only at higher temporal aggregation levels. Moreover, the values of the features may change considerably across different frequencies \citep{KOURENTZES2014291}. To that end, we expand the initial pool of features $\mathcal{F}$, originally computed for daily sales data, by performing temporal aggregation and calculating their values on a weekly and monthly level as well. In this regard, the final pool of features examined consists of a total of $3 \cdot P = 2508$ features per series.

\subsection{Feature selection}
After removing those with missing or unique values, 1654 features are kept.
Given the large number of features, which greatly increases the dimensionality and complexity of the analysis,  we employ a feature selection procedure by tailoring the approach proposed by \citet{lubba2019catch22} to meet our requirements.  The feature selection procedure involves three steps, namely statistical pre-filtering, performance evaluation, and redundancy minimization, as described in the following subsections. 

\subsubsection{Statistical pre-filtering}

The statistical pre-filtering step aims to remove the non-significant features rather than choosing the most meaningful ones. It involves the $z$-score standardization of the features and eliminating those that display the same or similar values across different series. To that end, we first select the features that can effectively differentiate distinct classes in the M5 data set in a statistically significant manner, considering four different item classification tasks in terms of states (three classes), stores (ten classes), product categories (three classes), and product departments (seven classes).

For each feature, we perform the nonparametric Kruskal-Wallis hypothesis test~\citep{kruskal1952use} to each of the four classification tasks, thus obtaining four $p$-values per feature accordingly. Note that, in contrast to the Analysis of Variance (ANOVA) method, the Kruskal-Wallis hypothesis test does not assume normally distributed populations, also performing better when population asymmetries are present \citep{hecke2012power}. We combine the four $p$-values as one single overall test for each feature using the Fisher’s method \citep{fisher1925statistical}. According to Fisher's method, $k$ independent $p$-values can be combined into one statistic that follows a $\chi^2$ distribution with $2 \cdot k$ degrees of freedom:
\begin{equation}
    X = -2 \sum_{i=1}^{k} \ln{p_i} \sim \chi^2(2 \cdot k),
    \label{eq:pvalue}
\end{equation}
where $p_i$ denotes the $p$-value for the $i$\textsuperscript{th} ($i = 1, \cdots, k$) statistical test (i.e., the $i$\textsuperscript{th} classification task in our case) and $k$ is the number of tests. The combined $p$-value for each feature can be obtained according to Equation~\eqref{eq:pvalue}. Finally, we apply the Holm-Bonferroni method \citep{holm1979simple} to correct the newly constructed $p$-value and reduce the type I errors of the hypothesis tests for the specified significance level, which may be caused due to the large number of tests involved in the process. To do so, we first sort all the $p$-values in ascending order and obtain their ranks. Then, we compare the original $p$-values with the corrected values $p/(n+1-\text{rank}(p))$, where $n$ represents the total number of features. The comparison continues until some original $p$-value is larger than the corresponding corrected $p$-value. Eventually, the features with smaller original $p$-values than the corrected ones at the stop point are retained. In this step,  132 of 1654 features are dropped with a significance level of 0.05.

\subsubsection{Performance evaluation}
In order to evaluate the quality of the extracted features, we employ the Regressional ReliefF (RReliefF) algorithm \citep{robnik2003theoretical}, which is an extension of the ReliefF algorithm for regression problems \citep{kononenko1994estimating}. ReliefF is a robust filtering method used to select features in multi-class classification problems, with the basic idea of identifying feature differences between nearest instance pairs. Specifically, ReliefF calculates a score $W_F$ for each feature $F$ and performs feature selection accordingly. According to \citet{kononenko1994estimating}, the $W_F$ scores derived by ReliefF are approximations of the following difference of probabilities:
\begin{equation}
\begin{aligned}
W_F = &p(\text{different~values~of}~F|\text{nearest~instances~from~a~different~class}) -\\ &p(\text{different~values~of}~F|\text{nearest~instances~from~the~same~class}).
\end{aligned}
\label{eq:wf0}
\end{equation}
Similarly, RReliefF calculates the probability of the predictions of two instances being different from each other. Based on Bayes' rule and Equation~\eqref{eq:wf0}, $W_F$ can be computed as follows
\begin{equation}
    W_F = \frac{p_{\text{diffP}|\text{diffF}} \cdot p_{\text{diffF}}}{p_{\text{diffP}}}-\frac{(1-p_{\text{diffP}|\text{diffF}}) \cdot p_{\text{diffF}}}{1-p_{\text{diffP}}}, 
\end{equation}
where
\begin{equation}
\begin{aligned}
    p_{\text{diffF}}= &p(\text{different~value~of}~F|\text{nearest~instances}),\\
    p_{\text{diffP}}= &p(\text{different~predictions}|\text{nearest~instances}),\\
    p_{\text{diffP}|\text{diffF}}= &p(\text{different~predictions}|\text{different~values~of}~F~\text{and~nearest~instances}).
\end{aligned}
\label{eq:wf-forecasting}
\end{equation}

In our case, seven point forecasts are used as predictions to evaluate the quality of the features more moderately, meaning that seven $W_F$ scores are computed for each feature. These scores are then combined into single quality vectors and used as input to the following step of the procedure to support further feature selection. For more details on the RReliefF algorithm and $W_F$ score estimation, please refer to the study of \citet{robnik2003theoretical}.

\subsubsection{Redundancy minimization}

We employ hierarchical clustering to reduce the redundancy in the obtained top-performing features using the Pearson correlation distance (cosine distance) with complete linkage at a threshold of 0.2~\citep{lubba2019catch22}, which makes the pairwise correlation coefficients of the features in the same cluster larger than 0.8, thus forming clusters of similarly performing features. To do so, we use the vectors generated from the previous step (performance evaluation) as input to the clustering method. In each cluster, the feature with the largest mean quality score is selected.

\subsubsection{Selected features}

After completing the feature selection procedure, we end up with a total of 42 features, in which 10 features are computed at a daily level, while 22 and 10 at a weekly and monthly level, respectively.  That reflects the benefits of considering temporal aggregation for extracting more meaningful and descriptive features. More details of the selected features $\mathcal{F}^*$ (including their names, description, packages employed, and temporal aggregation levels considered) are summarized in the supplementary material of the paper. It is evident that $\mathcal{F}^*$ involves a wide range of time series features, including information about the coefficients of the discrete Fourier transform, the location of the observations of the series, the variance and distribution of the data, the differential characteristics of the series, the entropy and linear trend of the series, the intermittency, and the statistics of popular time series analysis tests, among others. 

We should clarify that although some of the selected features may be challenging to interpret, our approach is still preferable over alternative ones that arbitrarily consider a limited sample of features, being more flexible and generic. This becomes evident if we consider that the selection of proper features depend strongly on the nature and the context of the application. Hence, automatic ways of feature selection have attracted much attention in the forecasting community~\citep[e.g.,][]{Li2020imaging}.

\section{Representativeness of the M5 competition data}

To evaluate the extent of the differences observed between the series of the M5 competition and those of other retail firms, we tried to retrieve data sets of retail companies that operate in different regions, consider different marketing strategies, and sell different types of products. Moreover, we tried to identify data sets that cover different operation periods than the one examined in M5 to account for possible changes in customer behavior, trends, and special events. 

To that end, we considered the data sets available in the online data science platform of Kaggle, used in past forecasting competitions \citep{BOJER2021587}. Although there is a large number of competition data sets available, most of them are not comparable to M5 as they involve series monitored at different temporal aggregation levels, refer to different industries and applications, or originate from the same retail firm (Walmart). This limited our choices to the data set used for conducting the ``Corporación Favorita Grocery Sales Forecasting'' competition, involving the daily unit sales of a major grocery retail firm in South America\footnote{The firm operates in a various countries of South America, including Ecuador, Colombia, Costa Rica, Chile, Panama, Paraguay, and Peru, but the data used cover stores located in Ecuador.}. The data (174,654 series) is provided at product-store level, as done in M5, and covers a period of about 1,114 days, ranging from January 2013 to August 2017, which overlaps to some extent with the period considered in M5. Moreover, the data set involves 16 states, 54 stores, and 33 product categories.

\begin{figure}[th]
\centering
    \includegraphics[width=\linewidth]{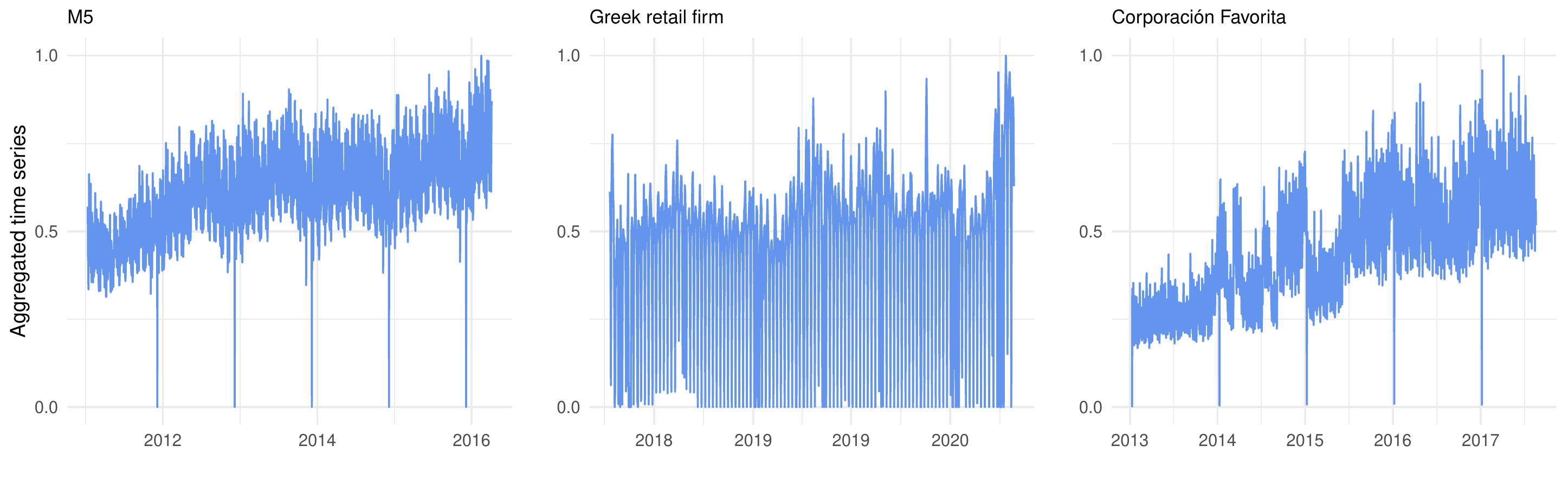}
    \caption{Aggregated unit sales of the examined data sets: M5 competition (Walmart), Greek supermarket chain, and Corporación Favorita. Sales are normalized using min-max scaling to facilitate comparisons.}
    \label{fig:seriesSum}
\end{figure}

In addition to the Corporación Favorita data set, we were able to acquire another one, generously provided to us by a major Greek supermarket chain, involving the daily unit sales of various products sold at 227 stores located all over Greece across 80 product categories. The data (7,248 series) is provided at warehouse level and covers a period of about 748 days, ranging from April 2018 to May 2020. Note that this period does not overlap with the one considered in M5, accounting also for the first wave of the COVID-19 outbreak and the respective lockdown effect. Note also that, similar to Corporación Favorita, the Greek retail firm drives its sales through promotions and discounts, in contrast to Walmart, which adopts a constant low-price marketing strategy. The aggregated unit sales of the examined data sets are presented in Figure \ref{fig:seriesSum}. Observe that the sales of the M5 and Corporación Favorita data set are characterized by greater trend compared to the Greek retailer, with those of Corporación Favorita being also more volatile. Moreover, the stores of the Greek retailer are closed on most Sundays, in contrast to those of the other two firms that are closed only on Christmas. However, as shown in Figure \ref{fig:seriesSum}, this is not the case in all weeks. Thus, Sundays were not removed from the Greek retailer's data and the selected time series features were calculated for all data sets in a consistent fashion, assuming a frequency of seven.

Apart from covering different regions, time periods, product categories, and marketing strategies, the series of the three data sets considered also display major differences in terms of intermittency (average inter-demand interval) and demand size erraticness \citep{SYNTETOS2005303}. As shown in Table \ref{tab:characteristics}, the majority of the M5 time series are intermittent, while most of the series of the Greek firm and Corporación Favorita are lumpy. However, in the latter case, the population of the four time series categories is far more balanced compared to the rest. These insights further motivate our research question, stressing the necessity to assess the representativeness of the M5 series in terms of feature spaces. Figure A of the supplementary material provides also useful visualizations regarding the distribution of the values of the selected features across the time series of the examined data sets.

\begin{table}[!h]
\caption{Overview of the examined data sets. For each data set, the region and period of operation are reported, along with the number of series involved and their average lengths. The percentages of erratic, lumpy, smooth, and intermittent series in each data set are also reported.}
\label{tab:characteristics}
\resizebox{\textwidth}{!}{\begin{tabular}{llccccccc}
\toprule
Data set & Region & Period &  Observations & Time series & Erratic & Lumpy & Smooth & Intermittent \\ \midrule 
M5 (Walmart) & USA & Jan 2011 - Apr 2016 & 1,507 & 30,490 & 2.88\% & 17.01\% & 6.76\% & 73.35\% \\
Greek retail firm & Greece & Apr 2018 - May 2020 & 748 & 7,248 & 18.10\% & 41.75\% & 10.58\% & 29.57\% \\
Corporación Favorita & Ecuador & Jan 2013 - Aug 2017 & 1114 & 174,654 & 20.65\% & 30.91\% & 23.07\% & 25.37\% \\
\bottomrule 
\end{tabular}}
\end{table}

\begin{figure}[th]
  \centering
  \begin{minipage}[b]{\textwidth}
  \centering
    \includegraphics[scale=0.45]{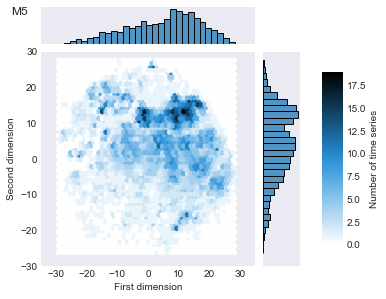}
  \end{minipage}
  \begin{minipage}[b]{0.49\textwidth}
  \centering
    \includegraphics[scale=0.45]{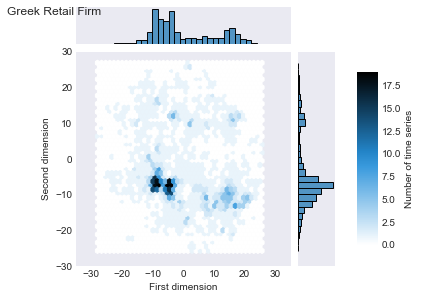}
  \end{minipage}
  \begin{minipage}[b]{0.49\textwidth}
  \centering
    \includegraphics[scale=0.45]{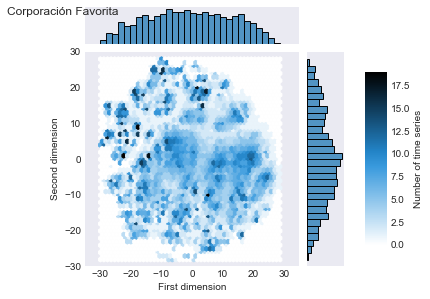}
  \end{minipage}
  \caption{The two-dimensional t-SNE instance spaces of the M5 (Walmart), Greek retail firm, and Corporación Favorita data sets.}
  \label{fig:Spaces}
\end{figure}

In order to visualize the discrepancies between the three data sets, we employ the approach proposed by \cite{KANG2020} which utilizes the t-Stochastic Neighbor Embedding~\citep[t-SNE,][]{van2008visualizing} to create feature spaces. Note that t-SNE is more effective than its linear counterparts (e.g., PCA, Principal Component Analysis) when considering multiple features that are correlated in a nonlinear fashion, placing similar data points close together while also keeping dissimilar ones far apart. While performing t-SNE, we use PCA to initialize the embedding, since informative initialization is more globally stable than random initialization in t-SNE~\citep{kobak2021initialization}. The generated feature spaces are presented in Figure \ref{fig:Spaces}. We observe that the series of the M5 competition successfully fill the overall instance space defined by the three retail firms, despite the fact that each data set displays different regions of higher densities. The same conclusion is drawn by observing the ranges and shapes of the distributions of the first and second t-SNE dimensions of the three data sets.

In order to quantify the differences and similarities observed among the three data sets, we proceed by computing the following measures: 

\begin{itemize}
  \item the miscoverage of data set A over data set B \citep{KANG2020}:
  \[ miscoverage_{A/B} = N^{-2} \displaystyle\sum_{i=1}^{N^{2}} (1-I_{i,A}) \times I_{i,B} ,\]
  where \(N^{2}=900\) is the number of squares of the grid superimposed on the overall space and  \( I_{i,A} \) equals to one if points in data set A fall within the \( i^{th} \) square, being  zero otherwise. An analogous definition is applied to \( I_{i,B} \) computed based on data set B.
  \item the percentage of points of data set A that lie within the space where data set B is not encountered (non-overlapping ratio):
  \[ NOR_{A/B} = \frac{\displaystyle\sum_{i=1}^{N^{2}} n_{i,A} \times (1-I_{i,B})}{\displaystyle\sum_{i=1}^{N^{2}} n_{i,A}} ,\]
  where \(n_{i,A}\) is the number of points of the data set A that fall within the \(i^{th}\) square.
\end{itemize}

These measures can effectively quantify both the spatial discrepancy of the feature spaces and the significance of that discrepancy in an intuitive way. Note however that this is true if we assume that the existence of a single series per square is adequate to indicate representativeness. The values of the two measures are summarized in Table \ref{tab:metrics}. As seen, although in 5.56\% of the feature space defined by Corporación Favorita and M5 there is at least one time series of the former data set but none of the latter, this part of space accounts for just 2.48\% of the overall series included in the Corporación Favorita data. Similarly, only 0.67\% of the space defined by the Greek retail firm and M5 is not represented by the latter, accounting for 0.37\% of the series of the Greek retailer. 

\begin{table}[!h]
\caption{Pairwise miscoverage and non-overlapping ratios among the data sets of the M5 competition (Walmart), the Greek retail firm, and Corporación Favorita.}
\label{tab:metrics}
\centering
\begin{tabular}{lcccc}
\hline
\multirow{2}{*}{Data set A} & \multicolumn{3}{c}{Data set B} \\
\cline{2-4}
& M5 & Greek retail firm & Corporación Favorita \\ 
\hline
&\multicolumn{3}{c}{Pairwise miscoverage}\\
\cline{2-4}
M5 (Walmart) & - & 0.67\% & 5.56\% \\
Greek retail firm & 18.78\% & - & 23.56\% \\
Corporación Favorita & 0.11\% & 0.00\% & - \\
\hline
&\multicolumn{3}{c}{Pairwise non-overlapping ratio}\\
\cline{2-4}
M5 (Walmart) & - & 14.70\% & 0.02\% \\
Greek retail firm & 0.37\% & - & 0.00\% \\
Corporación Favorita & 2.48\% & 24.23\% & - \\
\hline
\end{tabular}
\end{table}

Based on our results, we conclude that there are only small discrepancies between the examined data sets and that the findings of M5 can be relevant for various retail firms. Moreover, we find that although the size of the data set affects to some extent the degree of representativeness, M5 manages to effectively cover the feature space of the Corporación Favorita data, which consists of about 5.7 times more series and considerably more stores, locations, and product categories. Moreover, we find that the degree of miscoverage is not affected by the period examined, even when special events like the COVID-19 outbreak are present. This result reflects the diversity of the products and locations considered for constructing the M5 data set, as well as the representativeness of the period examined, accounting both for typical, business-as-usual days and special ones.   

To further validate our conclusions, we conduct a similar but simpler analysis that considers the six intuitive time series features proposed by \cite{Kang2017} (spectral entropy, strength of trend, strength of seasonality, seasonal period, first order autocorrelation, and optimal Box-Cox transformation parameter), plus two additional ones that can effectively capture intermittency (average inter-demand interval) and erraticness (squared coefficient of variation). Thus, we end up with eight primary features, each computed at a daily, weekly, and monthly level, resulting in a set of 24 features in total. After computing the values of these features for all the time series of the data sets, we apply a principal component analysis and use the first two components to project the series onto a 2-dimensional space to allow for interpretable data visualizations. The resulting graphs of this approach, shown in Figure B of the supplementary material of the paper, lead to similar findings with the previous approach in the sense that only small discrepancies can be identified between the three data sets examined.

\section{Conclusion}
In this study, we extended previous work done in feature-based time series instance space analysis, incorporating a plethora of features computed across multiple temporal aggregation levels and using an intuitive feature selection process to capture the key features of the M5 competition data set and evaluate the degree of its representativeness for the retail industry. To do so, we considered two additional retail sales data sets that cover different regions, time periods, product categories, and marketing strategies compared to M5. 

Our analysis has found no meaningful inconsistencies between the M5 data and the series of the two other grocery retailers, with any discrepancies observed between the examined data sets being minor. This was also true when data sets of significantly larger size and diversity in terms of stores, locations, and product categories were considered, as well as when different time periods were examined. However, we should note that our analysis is limited to companies that sell their products through brick-and-mortar stores, with a particular focus on groceries. As such, it could be the case that online retailers and firms that sell different types of products, such as pharmaceutical and technological ones, may experience different sales patterns. In addition, our analysis focuses on the sales data itself, thus ignoring the representativeness of the explanatory variables that typically accompany this data to enhance forecasting performance.

\section*{Acknowledgments}
\label{acknowledgements}

The authors are grateful to the editors and two anonymous
reviewers for their helpful comments that improved the contents of the paper. Yanfei Kang is supported by the National Key Research and Development Program (No. 2019YFB1404600) and the National Natural Science Foundation of China (No. 72021001 and No. 11701022).

\bibliographystyle{model5-names}  
\bibliography{main}

\newpage
\section*{Supplementary material}
\label{appendix:features}
{\scriptsize
\centering
    \begin{longtable}{lp{17em}p{24em}llll}
    \caption*{\textbf{Table A:} Description of the 42 time series features selected ($\mathcal{F}^*$) for conducting the analysis. D (daily), W (weekly), and M (monthly) indicate the temporal aggregation level at which the features are computed. The software package used for the estimation of the features is also provided.}\\
    \toprule 
        No. & Feature & Description & Package & D & W & M\\
    \midrule 
        1 & count\_below\_t\_0 & the percentage of observations in the series that are lower than or equal to zero & \pkg{tsfresh} & \checkmark & - & -\\
        2 & fft\_coefficient\_attr\_angle\_coeff\_63 & the angle of the 63\textsuperscript{rd} coefficient of the discrete Fourier transform & \pkg{tsfresh} & \checkmark & - & -\\
        3 & trough & trough & \pkg{tsfeatures} & \checkmark & - & -\\
        4 & fft\_coefficient\_attr\_angle\_coeff\_73 & the angle of the 73\textsuperscript{rd} coefficient of the discrete Fourier transform & \pkg{tsfresh} & \checkmark & - & -\\
        5 & has\_duplicate\_max & Boolean variable denoting whether the maximum value of the series is observed more than once & \pkg{tsfresh} &  \checkmark & - & -\\
        6 \& 7 & fft\_coefficient\_attr\_angle\_coeff\_1 & the angle of the 1\textsuperscript{st} coefficient of the discrete Fourier transform & \pkg{tsfresh} & \checkmark & - & \checkmark\\
        8 & fft\_coefficient\_attr\_angle\_coeff\_22 & the angle of the 22\textsuperscript{nd} coefficient of the discrete Fourier transform & \pkg{tsfresh} & \checkmark & - & -\\
        9 & fft\_coefficient\_attr\_angle\_coeff\_59 & the angle of the 59\textsuperscript{th} coefficient
        of the discrete Fourier transform & \pkg{tsfresh} & \checkmark & - & -\\
        10 & variance\_larger\_than\_standard\_de\-viation & Boolean variable denoting whether the variance of the series is greater than its standard deviation & \pkg{tsfresh} & \checkmark & - & -\\
        11 \& 12 & fft\_coefficient\_attr\_angle\_coeff\_26 & the angle of the 26\textsuperscript{th} coefficient of the discrete Fourier transform & \pkg{tsfresh} & \checkmark & \checkmark & -\\
        13 & augmented\_dickey\_fuller\_attr\_test\-stat\_autolag\_AIC  & the statistic of the ADF test where the lag is chosen by AIC & \pkg{tsfresh} & - & \checkmark & -\\
        14 & change\_quantiles\_mean\_isabs\_True\_ qh\_1.0\_ql\_0.8 & the mean, absolute value of consecutive changes of the series inside the corridor determined by the quantiles 0.8 and 1 of its distribution & \pkg{tsfresh} & - & \checkmark & -\\
        15 & fft\_coefficient\_attr\_imag\_coeff\_47 & the imaginary part of the 47\textsuperscript{th} coefficient of the discrete Fourier transform & \pkg{tsfresh} & - & \checkmark & -\\
        16 & fft\_coefficient\_attr\_real\_coeff\_36 & the real part of the 36\textsuperscript{th} coefficient of the discrete Fourier transform & \pkg{tsfresh} & - & \checkmark & -\\
        17 & number\_crossing\_m\_1 & the number of crossings of series on 1 & \pkg{tsfresh} & - & \checkmark & -\\
        18 \& 19 & fft\_coefficient\_attr\_angle\_coeff\_2 & the angle of the 2\textsuperscript{nd} coefficient of the discrete Fourier transform & \pkg{tsfresh} & - & \checkmark & \checkmark\\
        20 & fft\_coefficient\_attr\_angle\_coeff\_5& the angle of the 5\textsuperscript{th} coefficient of the discrete Fourier transform & \pkg{tsfresh} & - & \checkmark & -\\
        21 & fft\_coefficient\_attr\_imag\_coeff\_44 & the imaginary part of the 44\textsuperscript{th} coefficient of the discrete Fourier transform & \pkg{tsfresh} & - & \checkmark & -\\
        22 & cwt\_coefficients\_coeff\_12w\_2\_widths\_(2, 5, 10, 20) & the 12\textsuperscript{th} coefficient of the continuous wavelet transform for the Ricker wavelet & \pkg{tsfresh} & - & \checkmark & -\\
        23 & fft\_coefficient\_attr\_real\_coeff\_38 & the real part of the 38\textsuperscript{th} coefficient of the discrete Fourier transform & \pkg{tsfresh} & - & \checkmark & -\\
        24 & fft\_coefficient\_attr\_real\_coeff\_42 & the real part of the 42\textsuperscript{nd} coefficient of the discrete Fourier transform & \pkg{tsfresh} & - & \checkmark & -\\
        25 & fft\_coefficient\_attr\_angle\_coeff\_20 & the angle of the 20\textsuperscript{th} coefficient of the discrete Fourier transform & \pkg{tsfresh} & - & \checkmark & -\\
        26 & fft\_coefficient\_attr\_imag\_coeff\_49 & the imaginary part of the 49\textsuperscript{th} coefficient of the discrete Fourier transform & \pkg{tsfresh} & - & \checkmark & -\\
        27 & agg\_linear\_trend\_attr\_rvalue\_ chunk\_len\_10\_f\_agg\_var & the r value of a linear  regression with chunks for time series that were aggregated over chunks by variance & \pkg{tsfresh} & - & \checkmark & -\\
        28 & fft\_coefficient\_attr\_real\_coeff\_45 & the real part of the 45\textsuperscript{th} coefficient of the discrete Fourier transform & \pkg{tsfresh} & - & \checkmark & -\\
        29 & fft\_coefficient\_attr\_real\_coeff\_46 & the real part of the 46\textsuperscript{th} coefficient of the discrete Fourier transform & \pkg{tsfresh} & - & \checkmark & -\\
        30 & fft\_coefficient\_attr\_imag\_coeff\_46 & the imaginary part of the 46\textsuperscript{th} coefficient of the discrete Fourier transform & \pkg{tsfresh} & - & \checkmark & -\\
        31 & fft\_coefficient\_attr\_abs\_coeff\_49 & the absolute value of the 49\textsuperscript{th} coefficient of the discrete Fourier transform & \pkg{tsfresh} & - & \checkmark & -\\
        32 & approximate\_entropy\_m\_2\_r\_0.5 & approximate entropy & \pkg{tsfresh} & - & \checkmark & -\\
        33 & fft\_coefficient\_attr\_real\_coeff\_43 & the real part of the 43\textsuperscript{rd} coefficient of the discrete Fourier transform & \pkg{tsfresh} & - & \checkmark & -\\
        34 & fft\_coefficient\_attr\_real\_coeff\_48 & the real part of the 48\textsuperscript{rd} coefficient of the discrete Fourier transform & \pkg{tsfresh} & - & \checkmark & -\\
        35 & fourier\_entropy\_bins\_5 & fourier entropy considering bins of 5 observations & tsfresh  & - & - & \checkmark\\
        36 & change\_quantiles\_mean\_isabs\_True\_ qh\_0.4\_ql\_0.2 & the mean, absolute value of consecutive changes of the series inside the corridor determined by the quantiles 0.2 and 0.4 of its distribution & \pkg{tsfresh} & - & - & \checkmark\\
        37 & ratio\_beyond\_r\_sigma\_r\_1 & the ratio of values that are over $r \cdot \sigma$ away from the mean of the series & \pkg{tsfresh} & - & - & \checkmark\\
        38 & e\_acf1 & the first autocorrelation coefficient of the remainder of the STL decomposition & \pkg{tsfeatures} & - & - & \checkmark\\
        39 & change\_quantiles\_var\_isabs\_False\_qh\_ 0.4\_ql\_0.2 & the variance of consecutive changes of the series inside the corridor determined by the quantiles 0.2 and 0.4 of its distribution & \pkg{tsfresh} & - & - & \checkmark\\
        40 & change\_quantiles\_var\_isabs\_False\_qh\_ 0.6\_ql\_0.4 & the variance of consecutive changes of the series inside the corridor determined by the quantiles 0.4 and 0.6 of its distribution & \pkg{tsfresh} & - & - & \checkmark\\
        41 & large\_standard\_deviation\_r\_0.3 & Boolean variable denoting whether the deviation of the series is higher than 0.3 times the range of the series & \pkg{tsfresh} & - & - & \checkmark\\
        42 & agg\_linear\_trend\_attr\_rvalue\_chunk\_ len\_5\_ f\_agg\_max & the r value of a linear  regression with chunks for time series that were aggregated over chunks by maximum & \pkg{tsfresh} & - & - & \checkmark\\
        \bottomrule
    \end{longtable}
}

\newpage
\begin{figure}[th]
\label{fig:features}
\centering
\includegraphics[width=\linewidth]{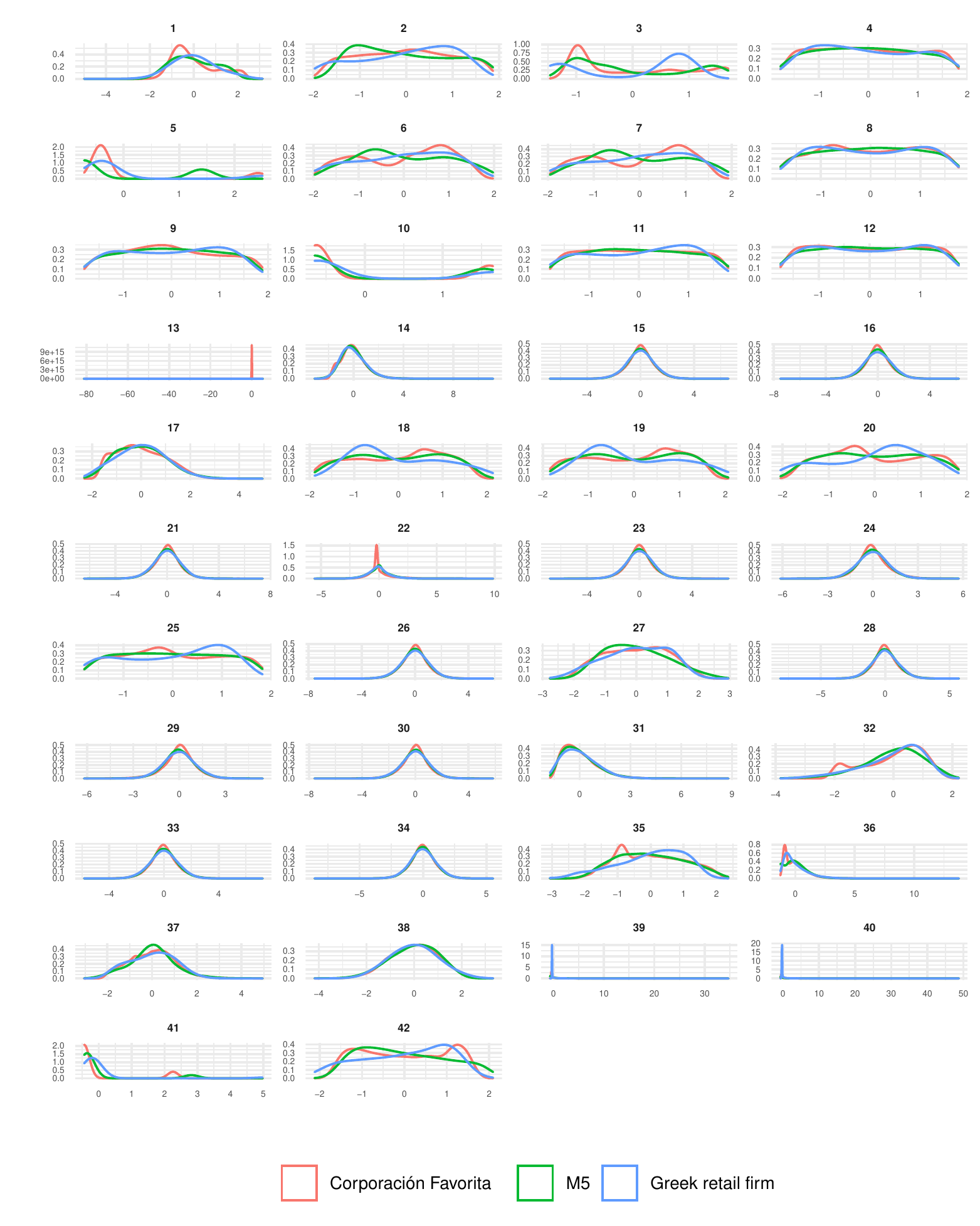}
    \caption*{\textbf{Figure A:} Distributions of the 42 time series features selected ($\mathcal{F}^*$) for conducting the analysis compared among the three examined data sets. The title of each plot indicates the number of each feature, as shown in Table A.}
\end{figure}

\newpage
\begin{figure}[th]
  \centering
  \begin{minipage}[b]{0.49\textwidth}
  \centering
    \includegraphics[scale=0.4]{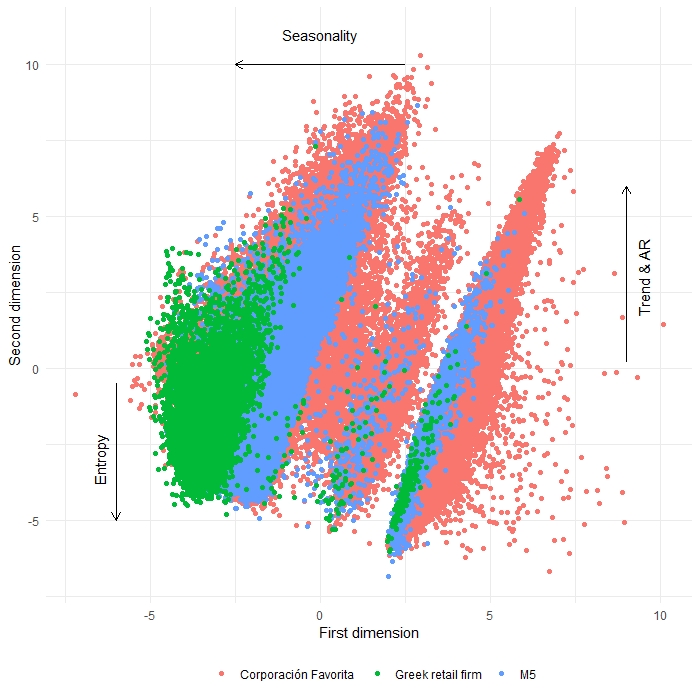}
  \end{minipage}
  \begin{minipage}[b]{0.49\textwidth}
  \centering
    \includegraphics[scale=0.45]{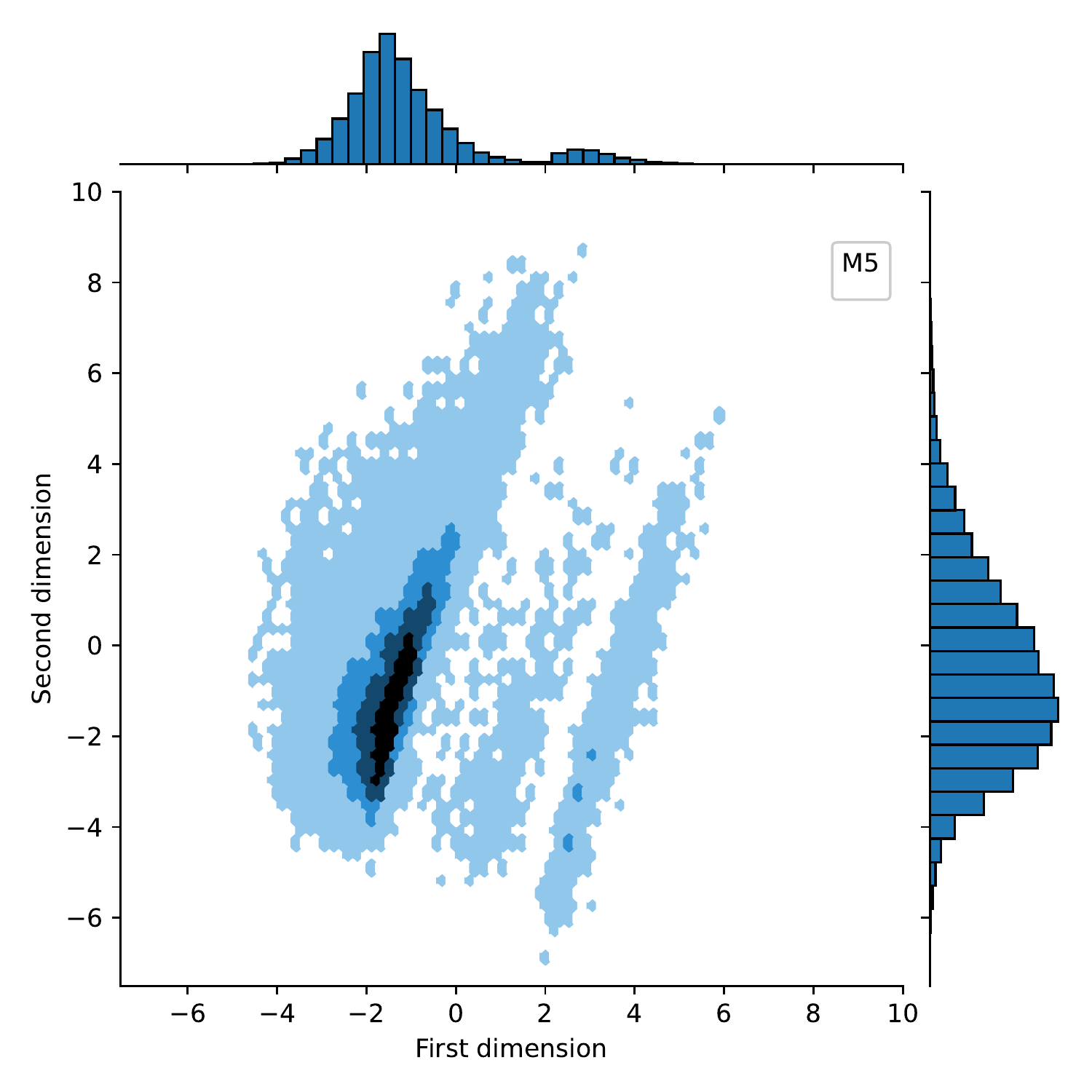}
  \end{minipage}
  \begin{minipage}[b]{0.49\textwidth}
  \centering
    \includegraphics[scale=0.45]{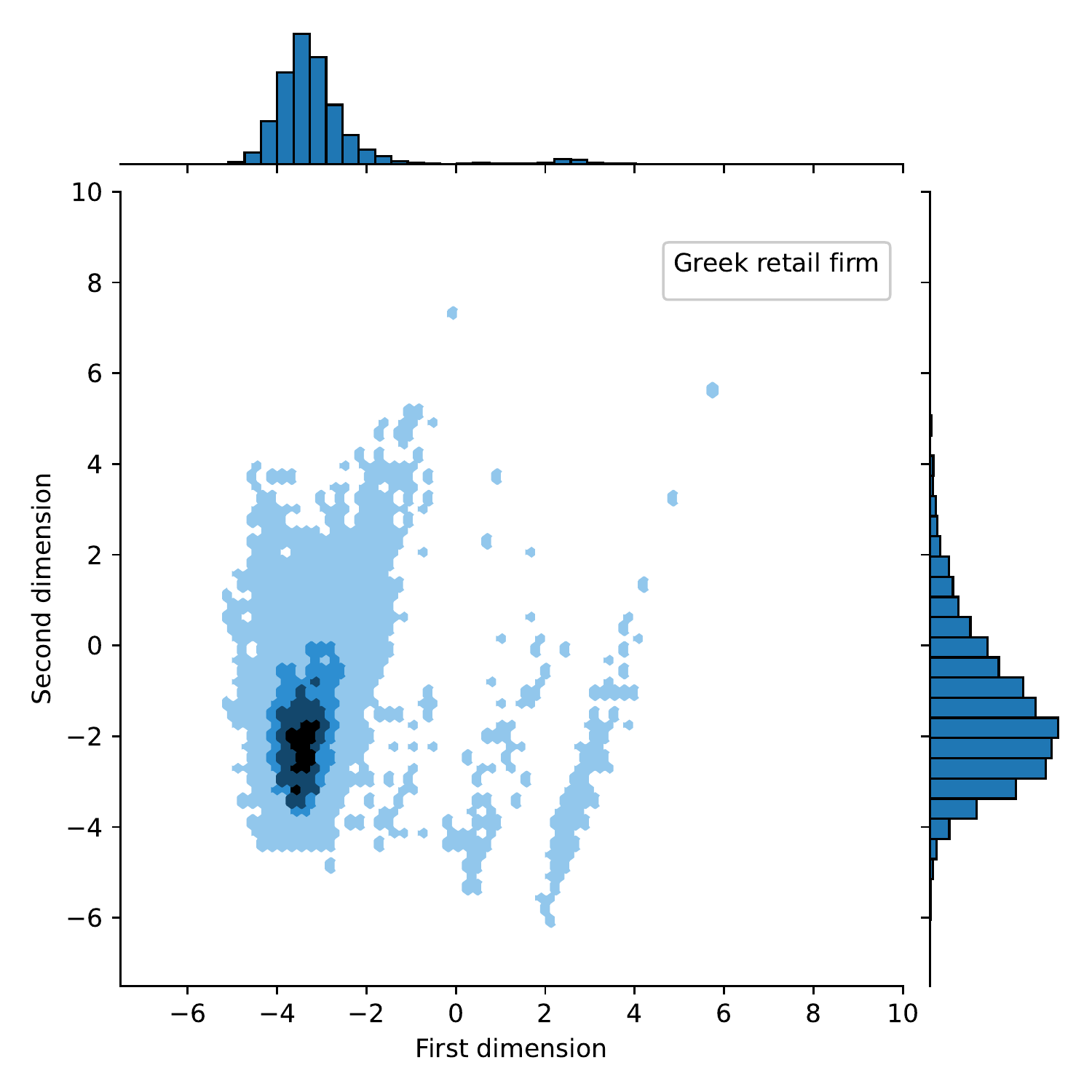}
  \end{minipage}
  \begin{minipage}[b]{0.49\textwidth}
  \centering
    \includegraphics[scale=0.45]{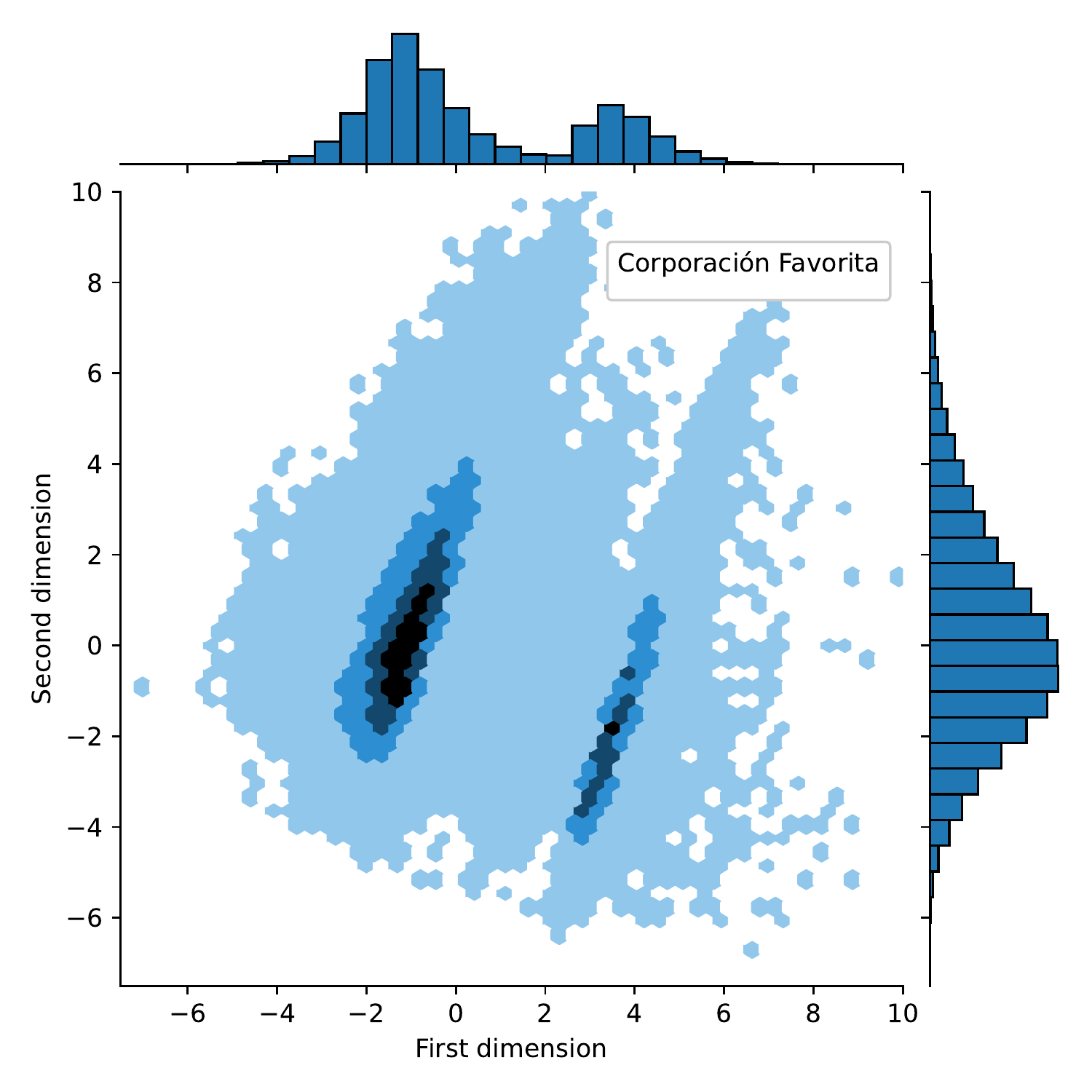}
  \end{minipage}
  \caption*{\textbf{Figure B:} Instance spaces of the M5 (Walmart), Greek retail firm, and Corporación Favorita data sets using the 6 series features proposed by \cite{Kang2017}, average inter-demand interval and squared coefficient of variation at daily, weekly and monthly levels.}
  \label{fig:PCASpaces}
\end{figure}

\end{document}